\documentclass[12pt]{iopart}
\usepackage{iopams}
\usepackage{graphicx}
\usepackage{color}
\usepackage[hidelinks]{hyperref}

\newcommand{\be}{\begin{equation}}
\newcommand{\ee}{\end{equation}}

\usepackage{xcolor}
\usepackage{ulem}

\begin{document}

\paper[Efficient gravitational-wave parameter estimation]{Efficient method for measuring the parameters encoded in a gravitational-wave signal}
\author{Carl-Johan Haster}
\address{School of Physics and Astronomy, University of Birmingham, Edgbaston, Birmingham B15 2TT, United Kingdom}
\author{Ilya Mandel}
\address{School of Physics and Astronomy, University of Birmingham, Edgbaston, Birmingham B15 2TT, United Kingdom}
\address{Monash Center for Astrophysics, Monash University, Clayton, VIC 3800, Australia}
\author{Will M. Farr}
\address{School of Physics and Astronomy, University of Birmingham, Edgbaston, Birmingham B15 2TT, United Kingdom}
\ead{cjhaster@star.sr.bham.ac.uk; imandel@star.sr.bham.ac.uk; w.farr@bham.ac.uk}

\begin{abstract}
Once upon a time, predictions for the accuracy of inference on gravitational-wave signals relied on computationally inexpensive but often inaccurate techniques. Recently, the approach has shifted to actual inference on noisy signals with complex stochastic Bayesian methods, at the expense of significant computational cost. Here, we argue that it is often possible to have the best of both worlds: a Bayesian approach that incorporates prior information and correctly marginalizes over uninteresting parameters, providing accurate posterior probability distribution functions, but carried out on a simple grid at a low computational cost, comparable to the inexpensive predictive techniques.
\end{abstract}

\submitto{\CQG}
\maketitle

\section{Introduction}
\label{sec:intro}

Advanced LIGO gravitational-wave (GW) detectors are expected to start science observations in 2015, with Advanced Virgo following shortly after \cite{AdvLIGO, AdvLIGO2, AdvVirgo,0264-9381-32-2-024001, scenarios}. A prime source predicted to be observable by these detectors are compact binary coalescences (CBCs) during which neutron star (NS) or black hole (BH) components of the binary are driven by GW emission through an inspiral into a merger. Binary neutron stars (BNS) have already been observed as pulsars in Galactic binary systems evolving due to emission of GWs \cite{2005ASPC..328...25W,Kramer06102006}; extrapolations from these observations yield estimated coalescence rates of BNS systems as detectable by the Advanced LIGO--Virgo network (\cite{2010CQGra..27q3001A} and references therein).

\subsection{Binary coalescence model}
\label{sec:BNS_ex}

The information available in an observed circular CBC event is encoded through the waveform $\vec{h}$, fully described by a 15-dimensional parameter vector $\vec{\theta}$ which in turn can further be divided into two parameter sets. The amplitude and phase evolution of the waveform are governed by the intrinsic parameters which include the masses and spins of the binary components. The extrinsic parameters describe the projection of the gravitational wave emitted by the binary onto the detectors and include the object's position and orientation on the sky, its luminosity distance $d_L$ as well as the time and phase of the waveform at coalescence ($t_c$ and $\phi_c$). The level of correlation between the two parameter sets is only marginal, apart for $t_c$ and $\phi_c$ which are strongly correlated with intrinsic parameters. Therefore we will simplify the analysis by fixing the remaining extrinsic parameters at their true values without significant impact on the recovery of the intrinsic parameters; we return to this point in \autoref{sec:conclusions}. 

In this study we consider as an example a BNS with component masses $m_1=1.45 M_{\odot}$ and $m_2=1.35 M_{\odot}$. While there is a strong degeneracy between the mass and spin parameters, which affects their measurement uncertainty \cite{2012PhRvD..86h4017B, F2Y_paper, 2014arXiv1411.6934B, 2014PhRvL.112j1101F, Hannam_et_al}, the BNS components are in this study taken as non-spinning for demonstration purposes, as our focus here is on parameter estimation techniques rather than astrophysical predictions. Therefore, the original 15-dimensional parameter space is reduced to only four parameters, the two masses as well as $t_c$ and $\phi_c$, without loss of generality in the method applied. The waveform approximant used is TaylorF2 \cite{PhysRevD.80.084043} which for a face-on binary directly overhead the detector describes the frequency-domain waveform as 
\be
\tilde{h}(f, \vec{\theta}) = \frac{1}{d_L}\sqrt{\frac{5}{24}}\pi^{-2/3}\mathcal{M}_c^{5/6}f^{-7/6}e^{i\Psi(f, \vec{\theta})} \, ,
\ee
assuming no higher-order corrections to the waveform amplitude. Here, $\mathcal{M}_c \equiv (m_1 m_2)^{3/5}(m_1 + m_2)^{-1/5}$ is the chirp mass and the phase $\Psi(f, \vec{\theta})$ is given as
\be
\Psi(f, \vec{\theta})= 2\pi f t_c - \phi_c - \frac{\pi}{4} + \frac{3}{128\eta\nu^5} \sum_{k=0}^7( \alpha_k +\beta_k\log(\nu))\nu^k
\ee
where $\nu \equiv (\pi (m_1 + m_2)f)^{1/3} $ and the coefficients $\alpha_k$, $\beta_k$ are functions of $\mathcal{M}_c$, the spins (here assumed to be zero) and the mass ratio $\eta \equiv (m_1 m_2)(m_1 + m_2)^{-2}$ for each post-Newtonian order $k/2$.

The likelihood of observing data $d \equiv h_T + n$, containing the true waveform $h_T$ and stationary, Gaussian instrumental noise $n$, given a waveform model $h(\vec\theta)$ described by parameters $\vec\theta$ is
\be
\label{eq:likelihood}
L(\vec{\theta})\propto\exp\left( -\frac{1}{2}\langle n+h_T-h(\vec{\theta})|n+h_T-h(\vec{\theta})\rangle\right).	
\ee
The term in the angle brackets is the noise-weighted power in the residuals after subtracting the assumed model from the data. The angle brackets denote an inner product between the Fourier transformed time series $\tilde{a}(f)$ and the complex conjugate of $ \tilde{b}(f)$ weighted by the noise power spectral density ${S_n(f)}$: 
\be
\langle a|b\rangle = 4 \, \mathrm{Re} \int_0^\infty \frac{\tilde{a}(f) \tilde{b}^\ast(f)}{S_n(f)}\,\mathrm{d}f.
\label{eq:inner_prod}
\ee
The likelihood \eref{eq:likelihood}, together with the specification for the waveform model and the noise power spectral density ${S_n(f)}$, which is chosen to match the Advanced LIGO zero-detuning high-power design sensitivity \cite{PSD:AL}, constitutes the model $H$.

For this study we will consider a BNS observed directly overhead in one detector at a signal to noise ratio $\rho=12$, defined as in \cite{2009PhRvD..79l2001A}:
\be
\label{eq:SNR}
\rho \equiv \max_{\vec{\theta}} \frac{\langle d | h(\vec{\theta}) \rangle}{\sqrt{\langle h(\vec{\theta}) | h(\vec{\theta}) \rangle}} \, .
\ee
The injected noise realisation in our example is chosen to be $n\equiv \vec{0}$, which fixes the maximum likelihood waveform to be $h_{T}$. 

\subsection{Bayesian inference}
Given the model $H$ as described in \autoref{sec:BNS_ex}, \eref{eq:likelihood} can be viewed as a probability density function (PDF) $p(d|\vec{\theta},H)$ for observing data $d \equiv h_T + n$ given parameters $\vec{\theta}$. Together with a PDF on the parameters detailing the prior knowledge and expectation about $\vec{\theta}$ under $H$, $p(\vec{\theta}|H)$, a posterior PDF can be obtained through Bayes' theorem as
\be
\label{eq:PDF}
p(\vec{\theta}|d,H)=\frac{p(\vec{\theta}|H)p(d|\vec{\theta},H)}{p(d|H)} \;.
\ee
The term in the denominator is called the evidence and can be used for comparing the effectiveness of different models' ability to describe the data; for this study, it is a normalizing constant for the posterior PDF.
The task of parameter estimation can therefore be defined as finding accurate and robust methods for evaluating the posterior probability given by \eref{eq:PDF} in order to obtain the accuracy with which different parameters can be recovered. This accuracy is usually quantified in terms of credible regions $\mathrm{CR}_p$ containing a fraction $p$ of the total posterior probability. These credible regions are defined as
\numparts
\be
\mathrm{CR}_p \equiv \min{A}
\ee
such that the parameter-space area $A$ satisfies
\be
\label{eq:cred_reg}
p = \int_A p(\vec{\theta}|d,H) \mathrm{d}\vec{\theta}\:.
\ee
\endnumparts
The posterior PDF will have the same dimensionality as $\vec{\theta}$ and a shape strongly dependent on the complexity of $H$. For the PDFs encountered in CBC data analysis, very intricate structures with high levels of multimodality and parameter degeneracies are common, and their prevalence has spurred the development and use of advanced analysis tools for sampling the posterior PDF.

\subsection{Stochastic sampling}
As part of the publicly available LSC Algorithm Library (LAL) \cite{LAL}, the Bayesian framework in \texttt{LALInference} implements several methods to stochastically traverse the parameter space \cite{lalinference, S6_PE}, creating a set of individual samples $\vec{\theta}_i$ distributed according to the posterior PDF in \eref{eq:PDF}. For this study, we used a parallel-tempered Markov chain Monte Carlo (MCMC) algorithm \cite{1953JChPh..21.1087M, HASTINGS01041970}, implemented in \texttt{LALInference} as \texttt{LALInferenceMCMC}.
The number of MCMC samples required to fully describe the posterior PDF scales only weakly with the number of dimensions, allowing for efficient exploration of high-dimensional parameter spaces. However, only a small fraction of the collected samples will be statistically independent, due to imperfect jump proposals in the complex multimodal parameter spaces. 
In order to accurately represent the posterior PDF for the 4-dimensional parameter vector $\vec{\theta}$ from \autoref{sec:BNS_ex}, $\mathcal{O}(10^3)$ independent samples are necessary \cite{2014PhRvD..89h4060S}. This requires a total of $\mathcal{O}(10^6)$ \texttt{LALInferenceMCMC} posterior samples, and therefore $\mathcal{O}(10^7)$ individual likelihood calculations spread across the 9 differently tempered parallel chains used in this example. 

\subsection{Paper organisation}
In \autoref{sec:Disc_eqPDF} we present the ``cumulative marginalized posterior'' method for accurately and efficiently evaluating credible region contours on a grid. We show that credible regions can be computed with a suitable choice of a low-density grid of samples, described in \autoref{sec:dens_grid}. This makes it possible to produce credible regions with the same accuracy as given by the stochastic samplers, but at a greatly reduced computational cost, as shown in \autoref{sec:key_results}. In \autoref{sec:alt_methods} we discuss alternative grid-based methods for estimating credible regions, including the iso-match contour technique advocated by Baird \textit{et al.} \cite{Baird_et_al}, and show their relative performance against the cumulative marginalized posterior method (\autoref{sec:comparasion}). We conclude in \autoref{sec:conclusions} with a discussion of these results and a proposal for implementing the cumulative marginalized posterior method in a low-latency parameter estimation framework for GW signals from CBCs.

\section{Discretizing the credible regions}
\label{sec:Disc_eqPDF}

As an alternative to the stochastic sampling methods we propose a method to evaluate the integral in \eref{eq:cred_reg} using samples at pre-determined coordinates in parameter space. The simplest implementation places samples into centres of pixels distributed in a uniform rectangular grid and approximates the posterior as constant within a pixel. The choice of the sampling grid is critical to the efficiency of this method, and will be discussed in \autoref{sec:dens_grid}. 

The required number of likelihood calculations, and therefore the required number of pixels, scales exponentially with the number of dimensions in the grid, so in order to minimize the computational requirements a modified version of \eref{eq:likelihood} was implemented. We use a likelihood function marginalized over $t_c$ and $\phi_c$, thus removing the need for sampling those dimensions without affecting the PDF in the mass parameters of interest. This marginalized likelihood function is presented in \cite{likelihood_tricks}, which also provides constraints on the resolution at which the waveforms must be sampled in both the frequency and time domains, in order to guarantee no loss of information when computing the inner products.

\subsection{Cumulative posterior on a grid}
\label{sec:Cum_pos_grid}
Taking advantage of the natural $[\mathcal{M}_c,\eta]$ parameterization of the waveforms, we constructed a uniformly spaced rectangular grid across these parameters. For each pixel in the grid, the likelihood corresponding to the value of $\vec{\theta}$ at its midpoint was calculated. We assumed a prior PDF on the mass distribution of BNS systems to be uniform in $[m_1,m_2]$ rather than $[\mathcal{M}_c,\eta]$ (transformed appropriately when working in the $[\mathcal{M}_c,\eta]$ space). The product of the likelihood and prior yield the numerator of \eref{eq:PDF}. The posterior PDF is obtained by normalizing this quantity by the evidence, the denominator in \eref{eq:PDF}, which is approximated as the sum of the likelihood-prior products over all pixels in the grid. The posterior PDF is assumed to be valid not just locally at $\vec{\theta}$ but instead across a whole pixel. Credible regions can then be defined by the set of pixels containing a fraction $p$ of the posterior PDF, accumulated when traversing the pixels in order of decreasing posterior values.

\subsection{Grid placement}
\label{sec:dens_grid}

To minimize computational cost and to ensure an accurate representation of the credible regions defined by the integral in \eref{eq:cred_reg}, the grid samples must be placed as sparsely as possible, subject to two constraints: (i) a sufficient fraction of the parameter space with significant posterior support is covered to enable accurate normalization of the posterior; and (ii) the error introduced by approximating the prior-likelihood integral over any pixel as the product of the prior and likelihood at the centre and the pixel area is within desired bounds. 

The required extent of the grid can be quantified in terms of the Mahalanobis distance $r(\vec{\theta})$ defined as
\be
\label{eq:MDist}
r = \sqrt{(\vec{\theta}-\vec{\mu})\Sigma^{-1}(\vec{\theta}-\vec{\mu})^T}
\ee
for a set of pixel coordinates $\vec{\theta}$ spanning a multivariate $N$-dimensional PDF $f(\vec{\theta})\equiv p(\vec{\theta}|d,H)$ with mean $\vec{\mu}$ and covariance $\Sigma$ \cite{MahalanobisDist}. When $f(\vec{\theta})$ is a bivariate Gaussian, 
the associated cumulative density function is given by $\Phi(r) = 1-e^{-r^2/2}$. Hence, for a maximum error $\epsilon = 1-\Phi(r)$ in the evidence contained within the grid, the grid must minimally contain the pixels bounded by a distance

\be
\label{eq:rBound}
r_b = \sqrt{-2\ln{\epsilon}}
\ee
away from the maximum (mean) of the PDF. The main purpose of this analysis is to construct credible regions whose $p\,$-value is known to an accuracy no worse than that of a stochastic sampler, which is of order $1\%$ for the $\mathcal{O}(10^3)$ samples we typically have (see \autoref{sec:key_results}). We therefore set $\epsilon=0.005$, and will correspondingly cover the region $r_b \le 3.25$ with a grid.

The minimum density of pixels within this bound is set by requiring that the approximate PDF, computed by discretely evaluating the posterior on a grid, is a sufficiently good approximation to the integral \eref{eq:cred_reg}. Here we use a very simple approximation, namely, we evaluate the integral as a Riemann sum over equal-sized pixels, setting the contribution of each pixel to the integral equal to the product of the pixel area and the value of the PDF at the centre of the pixel. In this case, a sufficient -- but not necessary -- condition for the total error on the PDF integral to be bounded by $\epsilon$ is for the fractional error in each pixel to be smaller than $\epsilon$. For an $N$-dimensional PDF $f(\vec{\theta})$, this fractional difference across a pixel centred at $\vec{\theta}_0$ is
\be
\label{eq:pixel_correction}
\left|\frac{f(\vec{\theta}_0)\Delta^N - \int^{\vec{\theta}_0 + \vec{\Delta}/2}_{\vec{\theta}_0 - \vec{\Delta}/2} \! f(\vec{\theta}) \, \mathrm{d}\vec{\theta}}{f(\vec{\theta}_0)\Delta^N}\right| \le \epsilon \, ,
\ee
where $\Delta$ is the pixel width. For a one-dimensional Gaussian distribution $f(\theta)$ with zero mean and variance $\sigma^2$, the integral in \eref{eq:pixel_correction} can be represented as a Taylor series: 
\be
\label{eq:Taylor_ser}
\eqalign{
\fl \int^{\theta_0 + \Delta/2}_{\theta_0 - \Delta/2} \! f(\theta) \, \mathrm{d}\theta \cr
\fl = \int^{\theta_0 + \Delta/2}_{\theta_0 - \Delta/2} \! f(\theta_0) \left( 1-\frac{\theta_0}{\sigma^2}(\theta-\theta_0) - \frac{1}{2\sigma^2}(\theta-\theta_0)^2 + \frac{\theta_0^2}{2\sigma^4}(\theta-\theta_0)^2 + \mathcal{O}((\theta-\theta_0)^3)\right) \, \mathrm{d}\theta \cr
\fl \approx f(\theta_0)\left( \Delta - \frac{\Delta^3}{24\sigma^2} + \frac{\theta_0^2\Delta^3}{24\sigma^4}\right)
}	
\ee
where the first non-zero correction term enters at the second order of the Taylor series since the approximated PDF is evaluated at the centre of the pixel. The integral will be dominated by the last term in \eref{eq:Taylor_ser} for increasing $|\theta_0|$; hence, the most stringent requirement on the pixel size will come from $\theta_0$ at the bounds of the integration domain. As discussed above, our analysis is restricted to $0 \le |\theta_0|/\sigma \le r_b$, so \eref{eq:pixel_correction} becomes
\be
\label{eq:pixel_req}
\left|\frac{(r_b^2-1)}{24}\frac{\Delta^2}{\sigma^2} \right| \le \epsilon\, ,
\ee
Hence, for $\epsilon=0.005$, the uniform grid size is $\Delta \approx 0.1 \sigma$, i.e., a total of $\sim 60$ pixels are required per parameter-space dimension. Therefore, assuming no correlations between parameters, a grid of $\approx 3500$ pixels is required to achieve $99.5\%$ coverage of the posterior region and sub-percent net credible region identification.

In practice, the grid size can be significantly reduced in a number of ways. The grid size need not be regular; rather than requiring a fixed fractional error per pixel, we could require a fixed contribution to the absolute error, which would allow us to increase the size of pixels near the bounds of the integration volume that contain a very small fraction of the PDF but set the most stringent requirements if the fractional error criterion is used. More accurate integration can be obtained by higher-order schemes, such as Simpson's rule, lowering the minimum number of grid points necessary to achieve a given accuracy. The grid need not be rectangular, but could be an $N$-dimensional sphere of dimensionless radius $r_b$, achieving a significant volume reduction in a high-dimensional space over a cube enclosing such a sphere, as assumed above. 

If the assumption of uncorrelated parameters is relaxed, \eref{eq:rBound} and \eref{eq:pixel_req} will still give the number of pixels required, but their relative placement needs to be altered. A misalignment between the grid and the PDF caused by correlated parameters or a non-ellipsoidal posterior PDF will reduce the validity of the previously given scaling relations. This can be overcome by either oversampling the grid; a coordinate rotation to align the grid and the PDF; or a dynamical placement of the pixels, adapting the local pixel density to a preliminary PDF estimated from a coarse grid across the parameter space. These extensions to the analysis will be investigated further in future work.

\subsection{Key results}
\label{sec:key_results}
\begin{figure}
\includegraphics[width=\textwidth,height=\textheight,keepaspectratio]{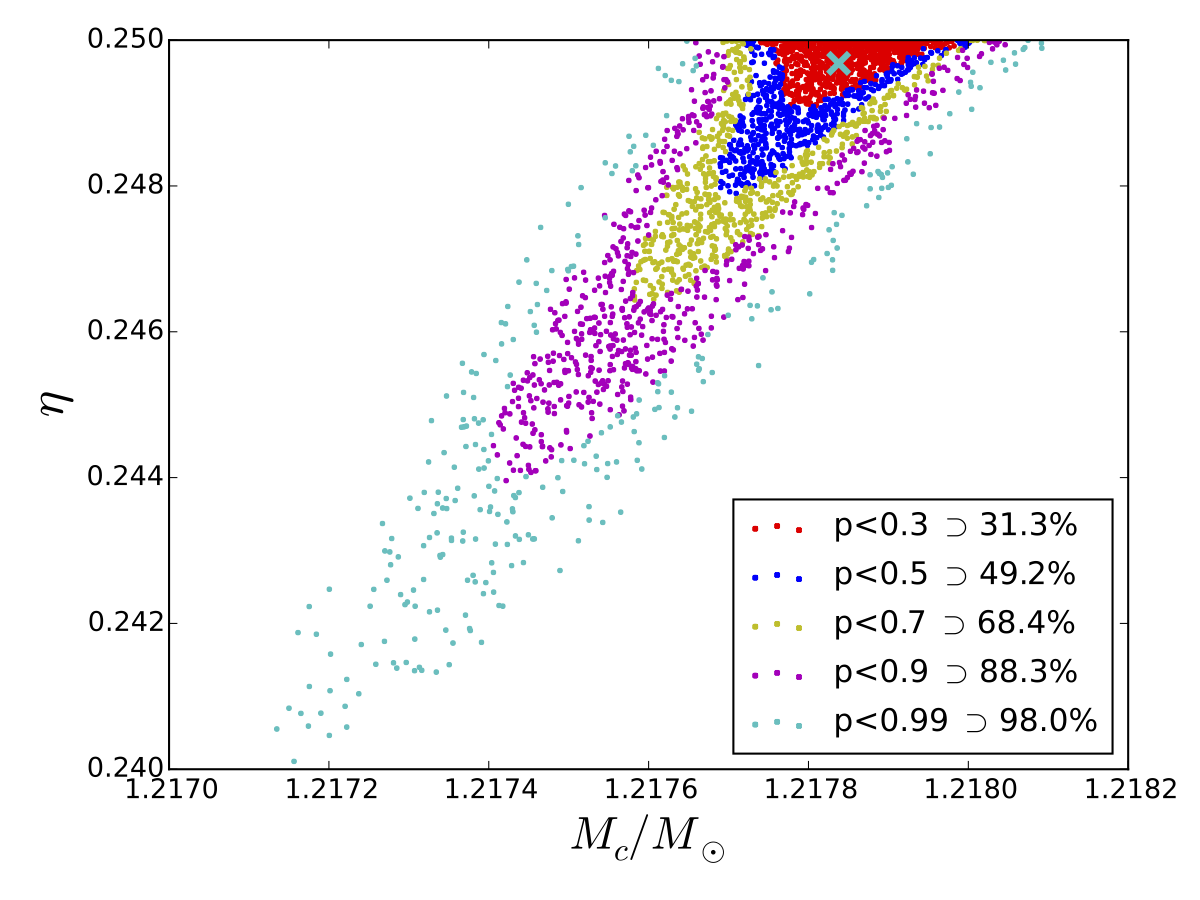}
\caption{MCMC samples (dots) in $[\mathcal{M}_c,\eta]$ space are colour-coded by the {\it cumulative marginalized posterior} credible region they fall into. The legend compares these credible regions against the fraction of MCMC samples falling into them, which corresponds to the stochastic estimate of the fraction of posterior contained within. The true parameters of the evaluated BNS system are shown at the turquoise cross (corresponding to $m_1 = 1.45 M_{\odot}$ and $m_2 = 1.35 M_{\odot}$).}
\label{fig:CumP_scatter} 
\end{figure}

For the example BNS system described in \autoref{sec:BNS_ex}, the grid-based cumulative marginalized posterior calculation can determine the credible regions of the posterior PDF to the same accuracy as \texttt{LALInferenceMCMC} using only a small fraction ($\sim 0.1\%$) of the computational cost of the stochastic sampler. 
This is illustrated in \autoref{fig:CumP_scatter}, which shows $2945$ independent samples of the posterior PDF produced by \texttt{LALInferenceMCMC}, colour coded by the credible region they fall into as given by the grid-based cumulative marginalized posterior. The credible regions are compared to the fraction of MCMC samples falling into the pixels contained within them, i.e., to the credible regions as estimated by the stochastic method. The number of MCMC samples falling within a claimed credible region is governed by a binomial distribution such that the uncertainty in the fraction of samples in $CR_p$ is $\sqrt{p(1-p)/N}$; e.g., for $N=2945$ and $p=0.3$, the uncertainty is $0.8\%$ -- consistent with the observed fluctuations in \autoref{fig:CumP_scatter}. 
The credible regions, and associated uncertainties, estimated for all $p\in[0,1]$ are shown in \autoref{fig:2D_pp_plot} as a complement to the discrete set of credible regions displayed here.

\section{Comparison with alternative methods: which approximations are warranted?}
\label{sec:alt_methods}

We have demonstrated that the cumulative marginalized posterior method is both accurate and computationally efficient with respect to stochastic samplers. We now explore which additional approximations can be made to simplify the analysis further; in the process, we investigate the validity of approximate techniques proposed by Baird \textit{et al.} and Hannam \textit{et al.} \cite{Baird_et_al,Hannam_et_al}. 

\subsection{Cumulative likelihood}
\label{sec:cum_L}
\begin{figure}
\includegraphics[width=\textwidth,height=\textheight,keepaspectratio]{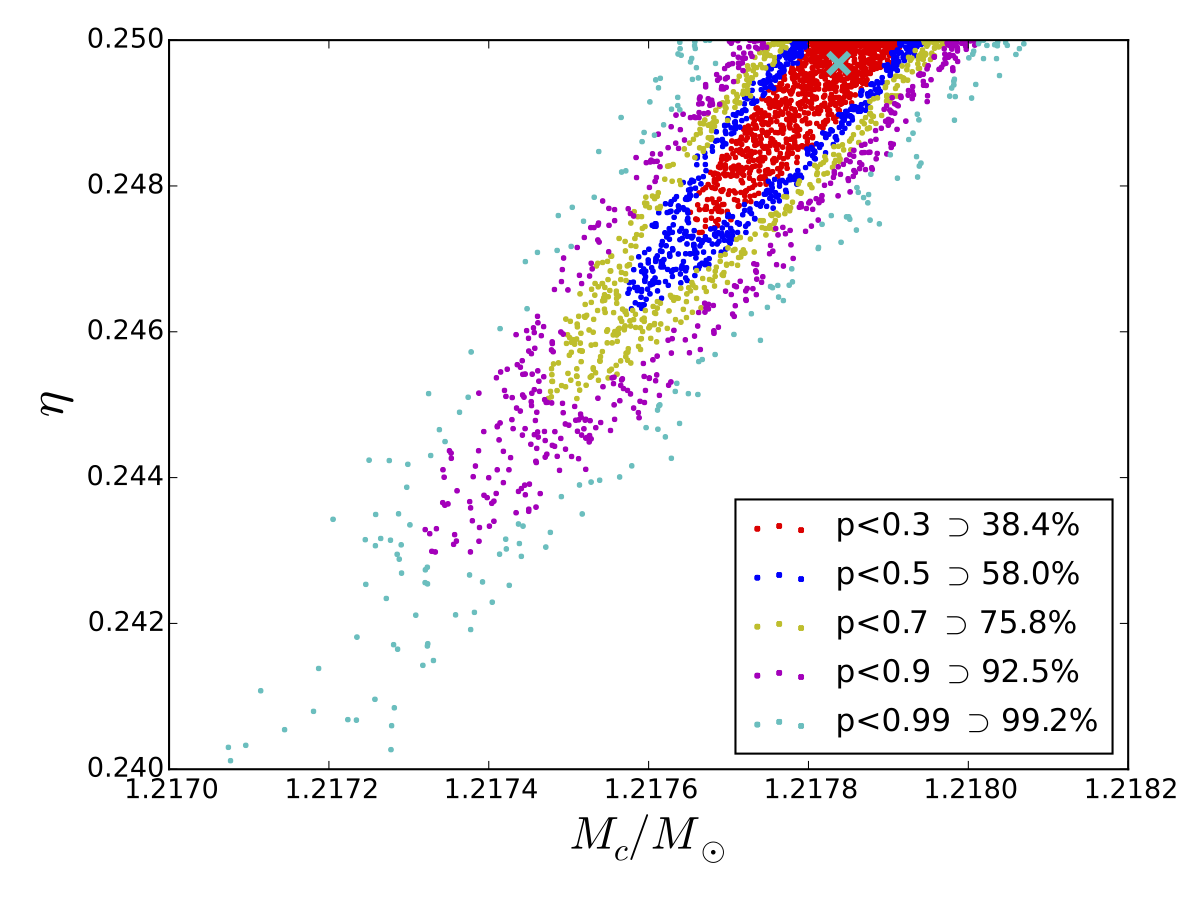}
\caption{MCMC samples (dots) in $[\mathcal{M}_c,\eta]$ space are colour-coded by the {\it cumulative maximized likelihood} credible region they fall into. The legend compares these credible regions against the fraction of MCMC samples falling into them, which corresponds to the stochastic estimate of the fraction of posterior contained within. The true parameters of the evaluated BNS system are shown at the turquoise cross (corresponding to $m_1 = 1.45 M_{\odot}$ and $m_2 = 1.35 M_{\odot}$).}
\label{fig:CumL_scatter} 
\end{figure}
In the Bayesian formalism used here, it is often assumed that the majority of the information about the posterior PDF originates from the likelihood function alone, with only a weak dependance on the prior PDF. To test this assumption, we repeated the analysis performed in \autoref{sec:Disc_eqPDF}, but without the inclusion of the prior detailed in \autoref{sec:Cum_pos_grid}. Evaluating the cumulative marginalized likelihood across the same $[\mathcal{M}_c,\eta]$ grid provided only negligible computational savings compared to the cumulative posterior as the overwhelming fraction of the computational cost is due to the likelihood calculations. The removal of the prior radically changed the shape of the credible regions from what was observed for the cumulative marginalized posterior in \autoref{fig:CumP_scatter} to the more ellipsoidal features shown in \autoref{fig:CumL_scatter}; moreover, the credible regions computed via the cumulative marginalized likelihood method no longer match the posterior PDF as evaluated with the MCMC sampler.

We next evaluated the effect of marginalization over $t_c$ and $\phi_c$ by replacing the previously used likelihood function with one which instead maximizes over $t_c$ and $\phi_c$, for both the cumulative posterior and likelihood methods. This carries greater computational savings compared to ignoring the prior, as maximizing the likelihood function reduces the computational time by $\sim 1/3$ compared to marginalizing over $t_c$ and $\phi_c$. However, replacing the correct marginalization with the computationally cheaper maximization produces credible regions which are no longer consistent with the posterior PDF estimated from MCMC calculations within the uncertainty discussed in \autoref{sec:key_results} (cf.~\autoref{fig:2D_pp_plot}).

While these two simplifications, particularly the use of maximization in lieu of marginalization, can yield reductions in computational complexity, the discrepancies introduced with respect to the credible regions produced by either \texttt{LALInferenceMCMC} or the cumulative marginalized posterior, are found to be outside the required tolerance level.

\subsection{Iso-Match contours and the Linear Signal Approximation}
\label{sec:Iso_match}

As an alternative approach for estimating credible regions and predicting parameter accuracy, Baird \textit{et al.} \cite{Baird_et_al} introduced a method, later implemented by Hannam \textit{et al.} \cite{Hannam_et_al}, based on the iso-match contours. This method relies on the validity of the Linear Signal Approximation (LSA) \cite{Vallisneri_Fisher, 2013PhRvD..88h4013R}. Under the LSA, waveforms are assumed to vary linearly with parameters, allowing a first-order expansion 
\be
h(\vec{\theta}) = h_T +h_i \Delta\theta^i \, ,
\ee
where $h_i$ is the partial derivative of the waveform with respect to the $i^{th}$ parameter and $ \Delta\theta^i = \theta^i - \theta_T^i$. When combined with \eref{eq:likelihood}, this yields the likelihood function
\be
L(\vec{\theta})\propto\exp\left( -\frac{1}{2}\langle h_i|h_j\rangle \Delta\theta^i\Delta\theta^j \right)\,,
\ee 
assuming $n\equiv \vec{0}$, expressed as a multivariate Gaussian centred at the true parameters $\vec{\theta_T}$ with covariance matrix ${\langle h_{i}|h_{j}\rangle}^{-1}$. 

The method of Baird \textit{et al.} uses the waveform match $M$ between waveforms $h_T$ and $h(\vec{\theta})$ defined as
\be
M = \max_{t_c, \phi_c} \frac{\langle h_T|h(\vec{\theta})\rangle}{\sqrt{\langle h_T|h_T\rangle\langle h(\vec{\theta})|h(\vec{\theta})\rangle}}\:,
\label{eq:match}
\ee
again maximizing over $t_c$ and $\phi_c$, as a proxy for the likelihood function. By assuming that the LSA is valid, Baird \textit{et al.} approximated credible region boundaries as contours of constant match,
\be
M_p = 1- \frac{\chi^2_N(1-p)}{2\rho^2} \, ,
\label{eq:matchth_apx}
\ee
via an $N$-dimensional $\chi^2$ distribution where $N$ is again the number of dimensions of the parameter space remaining after maximization. In addition to using maximization in lieu of marginalization, the validity of this approximation relies on the Gaussianity of the posterior, and does not include a priori information.

\begin{figure}
\includegraphics[width=\textwidth,height=\textheight,keepaspectratio]{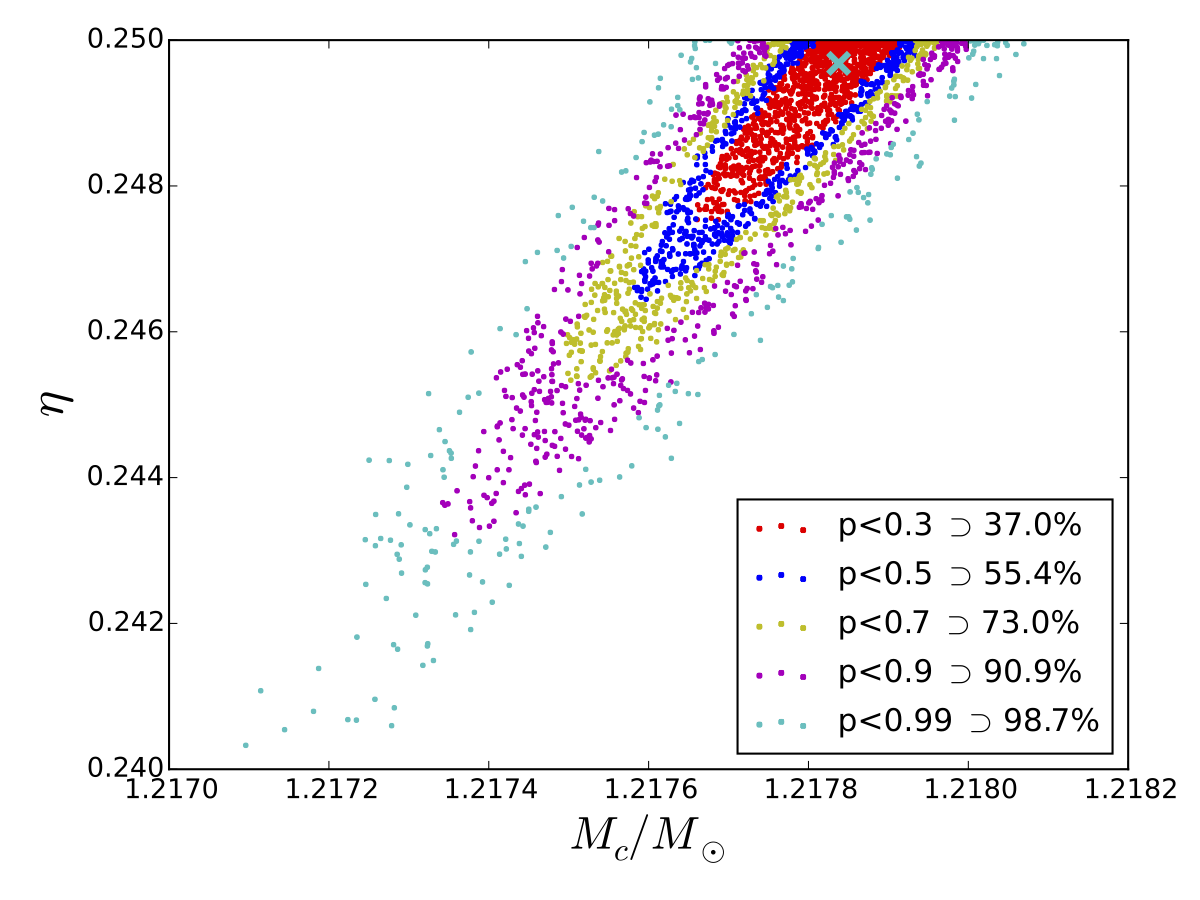}
\caption{MCMC samples (dots) in $[\mathcal{M}_c,\eta]$ space are colour-coded by the credible region they fall into as determined by the {\it iso-match contour} bounding them. The legend compares these credible regions against the fraction of MCMC samples falling into them, which corresponds to the stochastic estimate of the fraction of posterior contained within. The true parameters of the evaluated BNS system are shown at the turquoise cross (corresponding to $m_1 = 1.45 M_{\odot}$ and $m_2 = 1.35 M_{\odot}$).}
\label{fig:Baird_scatter} 
\end{figure}

Calculating the matches given for each pixel in the same $[\mathcal{M}_c,\eta]$ grid as used in \autoref{fig:CumP_scatter} and \ref{fig:CumL_scatter}, we defined credible regions as the pixels bounded by the iso-match contour in \eref{eq:matchth_apx}. 
\autoref{fig:Baird_scatter} compares credible regions given by iso-match contours against estimates from the fraction of MCMC samples falling within those contours; the differences between the two are statistically significant.
 
While the matches used by this method are calculated exactly, the restrictions implied by the LSA will lead the method to fail if the posterior PDF under investigation exhibits even a moderate level of non-Gaussianity. This can originate in the likelihood itself, or from the neglected contribution of the prior. This becomes clear for the BNS system evaluated here from the high degree of similarity between Figures \ref{fig:CumL_scatter} and \ref{fig:Baird_scatter}, indicating the validity of the LSA for this system. In this particular case, the Gaussianity of the likelihood in $[\mathcal{M}_c,\eta]$ space means that the posterior would have been Gaussian if the priors were flat in $[\mathcal{M}_c,\eta]$ space, so the method could have performed relatively well; it does not perform well for flat priors in $[m_1,m_2]$ space, as indicated by \autoref{fig:2D_pp_plot} (see below), because the posterior in this case is no longer Gaussian.

\subsection{Comparasion}
\label{sec:comparasion}

\begin{figure}
\includegraphics[width=\textwidth,height=\textheight,keepaspectratio]{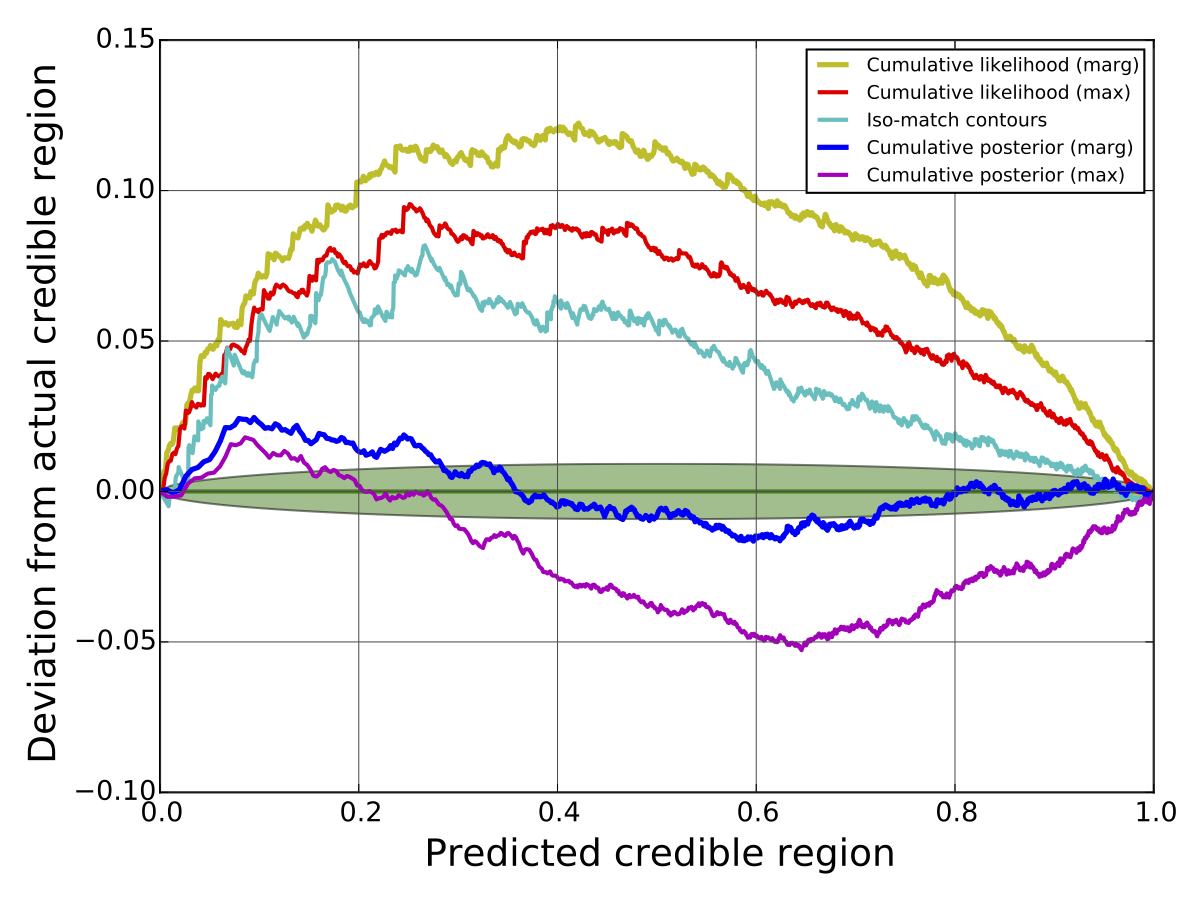}
\caption{Difference between the fraction of MCMC samples from the posterior PDF falling into credible regions predicted by the various methods described above and the expected credible level $p$, as a function of $p$. Continuous relations for $p \in [0,1]$ of the data in in \autoref{fig:CumP_scatter}, \ref{fig:CumL_scatter}
and \ref{fig:Baird_scatter} are shown by five curves, corresponding, from the top down, to the cumulative marginalized likelihood, cumulative maximized likelihood, iso-match contours, cumulative marginalized posterior, and cumulative maximized posterior. The filled ellipse indicates the expected 68\% level of fluctuation in the fraction of MCMC samples due to the finite number of samples.}
\label{fig:2D_pp_plot} 
\end{figure}

We compare all of the grid-based methods described above in \autoref{fig:2D_pp_plot}. We show the differences between the fraction of MCMC samples contained within the various methods for estimating credible regions corresponding to credible level $p$, and the value of $p$, for $p\in[0,1]$. Perfect agreement would correspond to a horizontal line at a deviation of zero. However, the finite number of stochastic MCMC samples from \texttt{LALInferenceMCMC} leads to statistical fluctuations in the deviation; their expected magnitude is indicated by an error ellipse (see \autoref{sec:key_results}) corresponding to one-$\sigma$ fluctuations.

The cumulative posterior method, using a likelihood function marginalized over $t_c$ and $\phi_c$, successfully estimates credible regions (apparent deviations at $p<0.3$ could be statistical, or may be due to the need for sub-pixel resolution to resolve such small credible regions). The comparison also further solidifies both the validity of the LSA for this system and the effect of the prior on the ability to recover consistent credible regions with respect to \texttt{LALInferenceMCMC}.

\section{Conclusions and future directions}
\label{sec:conclusions}

We have evaluated several grid-based methods for approximating the parameter credible regions for a CBC event, and compared these to regions estimated by the stochastic sampler \texttt{LALInferenceMCMC}. We found that evaluating the cumulative posterior on a relatively low-density grid allowed us to estimate credible regions to within the statistical uncertainty of the stochastic sampler at a small fraction of the computational cost ($\sim 0.1\%$), while marginalizing over the time and phase parameters and incorporating an arbitrary prior.

On the other hand, ignoring the prior or maximizing over $t_c$ and $\phi_c$ instead of marginalizing over them introduces a discrepancy in the recovered credible regions with respect to \texttt{LALInferenceMCMC}. In addition, we have demonstrated that the iso-match method proposed by Baird \textit{et al.} is overestimating the credible regions in $[\mathcal{M}_c ,\eta]$ space compared to a full Bayesian analysis.

\begin{figure}
\includegraphics[width=\textwidth,height=\textheight,keepaspectratio]{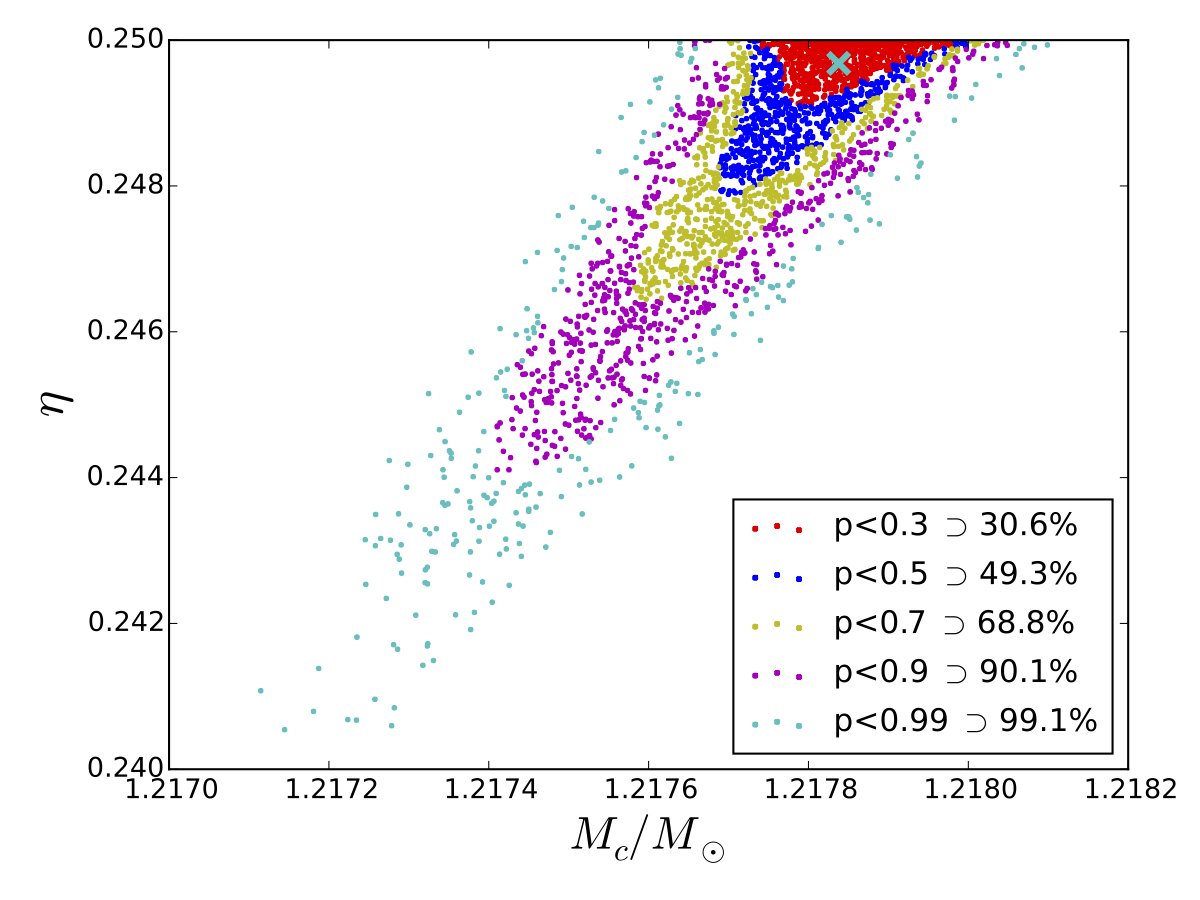}
\caption{MCMC samples (dots), now drawn from a 9-dimensional analysis, in $[\mathcal{M}_c,\eta]$ space are colour-coded by the {\it cumulative marginalized posterior} credible region they fall into, using the same two-dimensional grid as in previous figures. The legend compares these credible regions against the fraction of MCMC samples falling into them, which corresponds to the stochastic estimate of the fraction of posterior contained within, now allowing for effects of correlation against extrinsic parameters. The true parameters of the evaluated BNS system are shown at the turquoise cross (corresponding to $m_1 = 1.45 M_{\odot}$ and $m_2 = 1.35 M_{\odot}$).}
\label{fig:3IFO_vs_oneFixed} 
\end{figure}

The analysis has been performed on a binary observed in one detector at a fixed overhead and optimally oriented position and at a fixed distance giving $\rho=12$. Compared to an analysis of the same binary using a three-detector observation comprising data from the two LIGO observatories and the Virgo observatory (all operating at an the same sensitivity as assumed in \autoref{sec:BNS_ex}), where extrinsic parameters describing sky location, inclination, orientation and distance are allowed to vary, the recovered two-dimensional credible region in $[\mathcal{M}_c ,\eta]$ space, is not significantly altered (see \autoref{fig:3IFO_vs_oneFixed}). The analysis is implemented with the same $[\mathcal{M}_c ,\eta]$ grid as in previous figures, a grid designed for a two-dimensional analysis as described in \autoref{sec:dens_grid}. The discrepancy, if any, introduced by allowing for eventual correlation caused by the inclusion of extrinsic parameters is found to be within the statistical uncertainty from the limited number of samples from \texttt{LALInferenceMCMC} (c.f. \autoref{sec:key_results}).

The cumulative marginalized posterior method can easily be extended to include other parameters, especially spin \cite{Hannam_et_al}, but as the computational cost scales exponentially with the number of parameters, the cost quickly approaches and exceeds the computational requirements of the stochastic sampler implemented in \texttt{LALInference}. Our simple grid-based sampling implementation would be computationally competitive with the stochastic sampling for parameter spaces with up to {\it four} non-marginalized dimensions. However, the computational cost of the grid-based sampler could be further reduced through more efficient grid placement and more accurate integration algorithms. Moreover, while standard stochastic samplers such as \texttt{LALInferenceMCMC} are serial processes, all pixels in the sampling grid are completely independent, therefore trivially allowing for massive parallelization of the cumulative marginalized posterior analysis. 

Through these properties we envision the grid-based sampling method, using a cumulative marginalized posterior, to be implemented as a low-latency parameter estimation tool for the intrinsic parameters of a CBC candidate event, similar to the implementation of \textsc{bayestar} for the extrinsic parameters \cite{F2Y_paper,LeosThesis}. In practice, we won't know the true signal parameters which are needed for efficient and accurate grid placement. However, we can take the parameters of the highest-match template from the search pipelines used for the detection of CBC events \cite{ihope, Cannon2011Early}, which by design of the template banks will generally have $M>0.97$ \cite{1999PhRvD..60b2002O}, as the central point of the grid. The size of the grid can be initially estimated by comparing the SNRs reported by adjacent templates in the template bank. Subsequently, the grid can be adaptively refined: while we used a grid with uniformly spaced pixels for this study, the computational cost could be reduced further by implementing a non-uniform grid. The reduction in the absolute number of pixels required within the grid can enable the inclusion of additional non-marginalized dimensions while retaining the computational competitiveness against the stochastic sampling methods.

Moreover, predictive methods such as the iso-match method \cite{Baird_et_al,Hannam_et_al} or the effective Fisher matrix approach \cite{PhysRevD.87.024004,PhysRevD.89.064048} can be used in combination with the grid-based cumulative marginalized posterior technique to provide PDF estimates for dynamically laying out the grid. The cumulative marginalized posterior could also be implemented as a jump proposal for the intrinsic parameters as part of the stochastic samplers in \texttt{LALInference}. These possibilities will be explored further in future work. 

In addition, it is important to note that even though the method of parameter estimation with the grid-based sampling using a cumulative marginalized posterior has been presented here in the context of gravitational-wave astrophysics, the method itself is completely general and can be effective whenever the dimensionality of the parameter space is sufficiently small to make it competitive with stochastic samplers.

\ack
We are grateful for computational resources provided by the Leonard E Parker Center for Gravitation, Cosmology and Astrophysics at University of Wisconsin-Milwaukee. We received useful comments and suggestions from a number of colleagues at the University of Birmingham, including Christopher Berry, Walter Del Pozzo, Alberto Vecchio, and John Veitch. We are grateful to Richard O'Shaughnessy and Chris Pankow for careful reading of the manuscript. This work was partly supported by the Science and Technology Facilities Council and the Leverhulme Trust. IM acknowledges the hospitality of the Monash Center for Astrophysics, supported by a Monash Research Acceleration Grant (PI Y.~Levin). This is LIGO document number P1400251.\\

\section*{References}
\bibliographystyle{hunsrt} 
\bibliography{Efficient_Parameter_Estimation}

\end{document}